\documentstyle[12pt]{article}
\begin{document}

\baselineskip=23pt

\hspace{12cm} MPI-PhT 99-16

\hspace{12cm} April

\begin{center}
{\Large\bf Holography based on noncommutative geometry\\
and the AdS/CFT correspondence}

\bigskip
\bigskip

\bigskip

Zhe Chang

\medskip
{\em
Max-Planck-Institut f\"{u}r Physik\\
Werner-Heisenberg-Institut\\
F\"{o}hringer Ring 6, D-80805 M\"{u}nchen, Germany}

\medskip
and
\medskip

{\em
Institute of High Energy Physics, Academia Sinica\\
P.O.Box 918(4), 100039 Beijing, China}

\end{center}

\bigskip\bigskip

\bigskip

\centerline{\bf Abstract}
\medskip
Exponential regularization of orthogonal and Anti-de Sitter (AdS) space is presented based on
noncommutative geometry. We show that an adequately adopted noncommutative
deformation of geometry makes the holography of higher dimensional
quantum theory of gravity and lower dimensional theory possible.
Detail counting for observable degrees of freedom of quantum system of
gravity in the
bulk of noncommutative space $SO_q(3)$ and the classical limit of its
boundary surface $S^3$ is discussed. Taking noncommutivity effect into
account, we get the desired form of entropy for our world, which is consistent
with the physical phenomena associated with gravitational collapse.
Conformally invariant symmetry is obtained
for the equivalent theory of the quantum gravity living on the classical
limit of boundary of the noncommutative
AdS space. This is the bases of the AdS/CFT correspondence in string theory.

\bigskip\bigskip

\newpage

\section{Introduction}

By making use of the idea that a quantum system of gravity may possess far
less degrees of freedom than usually expected for a $3+1$ dimensional field
theory\cite{01}\cite{02}, it was
suggested recently that there is an AdS/CFT correspondence\cite{03}--
\cite{05}: the string theory (M theory) on background of the
form AdS$_d\times M_{D-d}$ is dual to a conformal
field theory living on the spacetime boundary. Here AdS$_d$ is an Anti-de
Sitter (AdS) space of spacetime dimension $d$, and $M_{D-d}$ is a certain
compactification space of dimension $D-d$ with $D=10$ for string theory
($D=11$ for M theory). A strong support for the
proposal comes from comparing spectra of Type IIB string theory on the
background of AdS$_5\times S^5$ and low-order correlation functions
of the $3+1$ dimensional ${\cal N}=4$ $SU(N)$ super Yang-Mills theory.
The dual super
Yang-Mills theory lives on the boundary of the AdS space. To
each field $\Phi_i$ there is a corresponding local operator ${\cal O}_i$
in the conformal field theory. The precise
relation between string theory in the bulk and field theory on the boundary is

\begin{equation}
Z_{\rm eff}(\Phi_i)=e^{iS_{\rm eff}(\Phi_i)}=\langle Te^{i\int_{\cal B}
\Phi_{b,i}{\cal O}^i}\rangle~,
\end{equation}
where $S_{\rm eff}$ is the effective action in the bulk and $\Phi_{b,i}$ is
the field $\Phi_i$ restricted to the boundary. In the large $N$ limit,
the string theory is weakly coupled and
supergravity is a good approximation to it. Thus it is possible to describe
a precise recipe expressing correlation functions of the ${\cal N}=4$
super Yang-Mills theory in four dimensions in
terms of a calculations of tree approximation to supergravity performed
in the bulk.

At first sight, it seems very strange for us that quantum theories in different spacetime
dimensions ever could be equivalent in any sense. The key to understand it is
the fact that the theory in the
larger dimension is always a quantum theory of gravity. For such theories the
concepts of holography has been introduced by 't Hooft\cite{01} based on
phenomenological study of the black hole theory and was referred as a
generic properties of quantum system of gravity.  The
AdS/CFT correspondence is just an example of the realization of the
holography on quantum theory of gravity. From point of view of general
relativity, gravity is nothing but spacetime geometry. To check
the Maldacena conjecture, one has to start from investigating of
adequate geometry description for quantum theory of gravity. Because the
holography is a generic properties of the quantum gravity, in principle,
it should be deduced naturally from  geometric properties of spacetime.
't Hooft showed that simple regularization of spacetime can not
give correct account of observable degrees of freedom for quantum theory of gravity.
On the other hand, the poor understanding of physics at Planck scale also
indicates that the small scale structure of spacetime might not be
adequately described by classical continuum geometry. Thus, new geometry
should be introduced for quantum gravity.
It has long time been suspected that the noncommutitive spacetime
might be a realistic picture of how spacetime behaves near
the Planck scale\cite{06}\cite{07}.  Strong quantum fluctuations of gravity may make points
in space fuzzy.
The noncommutative geometry description\cite{08}\cite{09} is a strong candidate for
quantum theory of gravity. We wish the holography can be obtained explicitly
based on noncommutative geometry picture of quantum gravity and show the
AdS/CFT correspondence by demonstrating conformally invariant symmetry on
the boundary surface of noncommutative AdS space.

In this paper, we present a kind of special regularization with exponentially increasing
spacetime cutoff for both orthogonal and AdS space based on noncommutative
geometry. It is used to argue that the same minimal cutoff of any geometry
is the Planck scale $l_p$, that the most direct and obvious physical cutoff is
from the formation of microscopic black holes as soon as too much energy would be accumulated
into too small a region\cite{01}. We show that an adequately adopted noncommutative
deformation of geometry makes the holography of higher dimensional
quantum system of gravity and lower dimensional theory possible.
As an example, detail calculations are carried out for the counting of
observable degrees of freedom of quantum gravity in the
bulk of noncommutative space $SO_q(3)$ and the classical limit of its
boundary surface $S^3$. Results show that a very small (may be $<10^{-15}$)
displacement of the noncommutative deformation parameter from its classical
value $1$ reduces sharply the entropy of the quantum system of gravity.
The desired entropy expression $S=4\pi M^2 +C$ of the universe can be
deduced naturally. Conformally invariant symmetry is
obtained for the equivalent theory of the quantum gravity, which lives on
the boundary of the noncommutative
AdS space. This is the bases of the AdS /CFT correspondence in
nonperturbative string theory and M theory.

This paper is organized as follows. In Section 2, we discuss the noncommutative
orthogonal space $SO_q(3)$. The algebra satisfied by the coordinates and
derivatives is decoupled into three independent subalgebras by introducing of
a new set of variables. Quantum coherent states are constructed as reference
ones for investigating representations of the spacetime algebra. Forms
of Hilbert space show clearly that the noncommutative spacetime is discretely
latticed
with exponentially increasing space distances. The noncommutative deformation parameter is
determined by an algebraic equation. The noncommutative space $SO_q(3)$ has a same
entropy or observable degrees of freedom with the classical
$S^3$ surface. Section 3 is devoted to the study of noncommutative AdS$_{2n}$
space. Conjugate operation is set up for the noncommutative AdS space. This
conjugation has a induced counterpart for the set of decoupled coordinates
and derivatives. Hilbert space of the noncommutative AdS$_{2n}$ space is
constructed based on quantum coherent states. Discrete lattice structure of the noncommutative AdS$_{2n}$ with
exponentially increasing space distances is obtained. Holography makes the
quantum system of gravity on the noncommutative AdS$_{2n}$ space equivalent
to the conformally invariant quantum theory living
on the classical limit of its boundary surface. This is crucial for the
AdS/CFT correspondence of string theory and M theory. By almost same
procedures, properties
of the noncommutative AdS$_{2n-1}$ space are shown in Section 4. In Section 5,
some concluding remarks are given.

\section{Noncommutative geometry and holography}

We begin by discussing the quantum space $SO_q(3)$ with coordinates $x^i$
($i=-,~0,~+$). The commutation relations\cite{10} among coordinates $x^i$ are

\begin{equation}
\begin{array}{l}
x^-x^0=qx^0x^-~,~~~~~x^0x^+=qx^+x^0~,\\
x^{+}x^{-}-x^{-}x^{+}=(q^{1/2}-q^{-1/2})x^0x^0~.
\end{array}
\end{equation}
The algebra satisfied by derivatives is of the form

\begin{equation}\label{diff}
\begin{array}{l}
\partial_-\partial_0=q^{-1}\partial_0\partial_-~,~~~~~
\partial_0\partial_+=q^{-1}\partial_+\partial_0~,\\
\partial_-\partial_+-\partial_+\partial_-=(q^{1/2}-q^{-1/2})\partial_0
\partial_0~.
\end{array}
\end{equation}
The action of derivatives on the coordinates is

\begin{equation}
\begin{array}{l}
\partial_-x^-=1+q^2x^-\partial_-+\lambda q x^0\partial_0
+\lambda (q-1) x^+\partial_+~,\\
\partial_-x^0=qx^0\partial_--q^{1/2}\lambda x^+\partial_0~,~~~~~
\partial_-x^+=x^+\partial_-~,\\
\partial_0x^-=qx^-\partial_0-q^{1/2}\lambda x^0\partial_+~,\\
\partial_0x^0=1+qx^0\partial_0+q\lambda x^+\partial_+~,~~~~~
\partial_0x^+=qx^+\partial_0~,\\
\partial_+x^-=x^-\partial_+~,~~~~~\partial_+x^0=qx^0\partial_+~,~~~~~
\partial_+x^+=1+q^2x^+\partial_+~.
\end{array}
\end{equation}
It is convenient to introduce the dilatation operators

\begin{equation}
\begin{array}{l}
\mu_+=1+q\lambda x^+\partial_+ ~,\\
\Lambda=1+q\lambda\displaystyle\sum_{j=0,\pm}x^j\partial_j+q^3\lambda^2
\left(q^{-1/2}x^-x^++\frac{q}{1+q}x^0x^0\right)\left(q^{-1/2}\partial_+
\partial_-+\frac{q}{1+q}\partial_0\partial_0\right)~.
\end{array}
\end{equation}
The dilatation operators obey

\begin{equation}
\begin{array}{l}
\mu_+ x^+=q^2x^+\mu_+~,~~~~~\mu_+\partial_+=q^{-2}\partial_+
\mu_+~,\\
\Lambda x^k=q^2 x^k \Lambda~,~~~~~\Lambda\partial_k=q^{-2}\partial_k\Lambda~,
~~~{\rm for}~k=0,\pm~.
\end{array}
\end{equation}
The real form $SO_q(3,R)$ (noted simply as $SO_q(3)$ whenever no confusion
rised) of the noncommutative space $SO_q(3)$ is obtained by a consistent
conjugation

\begin{equation}\label{conjugation}
\begin{array}{l}
\overline{x^i}=C_{ji}x^j~,\\
\overline{\partial_i}=-q^{-2}C_{ij}\Lambda^{-1}\displaystyle\left(q^{-1/2}
[\partial_+\partial_-,x^j]+\frac{q}{1+q}[\partial_0\partial_0,x^j]\right)~,
\end{array}
\end{equation}
where the metric $C_{ij}$ is of the form
$$C=\left(\begin{array}{ccc}
           & &q^{-1/2}\\
           &1& \\
         q^{1/2}& &\end{array}\right)~.$$
It is should be noticed that throughout this paper we limit us at the case of
$q$ being real.

By making use of the dilatation operators $\mu_+$ and $\Lambda$, we introduce a new set
of coordinates and derivatives

\begin{equation}
\begin{array}{l}
{\cal X}^-=\Lambda^{-1/2}\mu_+^{-1/2}\displaystyle\left(x^-+q^{3/2}\lambda
\left(q^{-1/2}x^-x^++\frac{q}{1+q}x^0x^0\right)\partial_+\right)~,\\
{\cal D}_-=q^{-1}\Lambda^{-1/2}\mu_+^{-1/2}\displaystyle\left(\partial_-+q^{3/2}\lambda
\left(q^{-1/2}\partial_+\partial_-+\frac{q}{1+q}\partial_0\partial_0\right)
x^+\right)~,\\
{\cal X}^0=\mu_+^{-1/2}x^0~,~~~~~{\cal D}_0=\mu_+^{-1/2}\partial_0~,\\
{\cal X}^+=x^+~,~~~~~{\cal D}_+=\partial_+~.
\end{array}
\end{equation}
In terms of the new variables,  the commutation relations among
coordinates and derivatives of the noncommutative space $SO_q(3)$ are transformed as

\begin{equation}
\begin{array}{l}
{\cal D}_-{\cal X}^--{\cal X}^-{\cal D}_-=\mu_-^{-1}~,~~~~~~
\mu_- {\cal X}^-=q^2{\cal X}^-\mu_-~,\\
\mu_-{\cal D}_-=q^{-2}{\cal D}_-\mu_-~,~~~~~~\mu_-^{-1}\equiv
1+(q^{-2}-1){\cal X}^-{\cal D}_-~;\\
{\cal D}_0{\cal X}^0-{\cal X}^0{\cal D}_0=\mu_0^{1/2}~,~~~~~~
\mu_0 {\cal X}^0=q^2{\cal X}^0\mu_0~,\\
\mu_0{\cal D}_0=q^{-2}{\cal D}_0\mu_0~,~~~~~~\mu_0^{1/2}\equiv
1+(q-1){\cal X}^0{\cal D}_0~;\\
{\cal D}_+{\cal X}^+-{\cal X}^+{\cal D}_+=\mu_+~,~~~~~~
\mu_+ {\cal X}^+=q^2{\cal X}^+\mu_+~,\\
\mu_+{\cal D}_+=q^{-2}{\cal D}_+\mu_+~,~~~~~~\mu_+^{1/2}=
1+(q^2-1){\cal X}^+{\cal D}_+~;\\[1mm]
[{\cal X}^i,{\cal X}^j]=0~,~~~~[{\cal D}_i,{\cal D}_j]=0~,~~~~
[{\cal D}_i,{\cal X}_j]=0~,\\[1mm]
[\mu_i,{\cal X}^j]=0~,~~~~[\mu_i,{\cal D}_j]=0~,~~~~~~{\rm for}~
i\not= j~.
\end{array}
\end{equation}
The noncommutative surface $S_q^3$  in terms of the set of independent
operators ${\cal X}^j$ and ${\cal D}_j$ is of the form

\begin{equation}
\frac{q^{-1}}{1+q}{\cal X}^0{\cal X}^0+q^{-1/2}\Lambda^{1/2}\mu_+^{-1/2}
{\cal X}^+{\cal X}^-=R^2~.
\end{equation}
At the $q\rightarrow 1$ limit, $S_q^3$ reduces to the familiar $S^3$ surface
with radius $R$.

The conjugate operation on ${\cal X}^j$ and ${\cal D}_j$ is induced by
what on $x^j$ and $\partial_j$ (Eq.(\ref{conjugation})),

\begin{equation}
\begin{array}{l}
{\overline{\cal X}^-}=\displaystyle\left({\overline x^-}+q^{3/2}\lambda
{\overline\partial_+}
\left(q^{-1/2}{\overline x^+}{\overline x^-}+\frac{q}{1+q}{\overline x^0}
{\overline x^0}\right)\right){\overline \mu_+}^{-1/2}
{\overline\Lambda}^{-1/2}~,\\
{\overline{\cal D}_-}=q^{-1}\displaystyle\left({\overline\partial_-}+
q^{3/2}\lambda {\overline x^+}
\left(q^{-1/2}{\overline\partial_-}{\overline\partial_+}-\frac{q}{1+q}
{\overline\partial_0}{\overline \partial_0}\right)
\right){\overline \mu_+}^{-1/2}{\overline \Lambda}^{-1/2}~,\\
{\overline {\cal X}^0}={\overline x^0}{\overline \mu_+}^{-1/2}~,~~~~~
{\overline {\cal D}_0}={\overline \partial_0}{\overline \mu_+}^{-1/2}~,\\
{\overline {\cal X}^+}={\overline x^+}~,~~~~~{\overline {\cal D}_+}=
{\overline \partial_+}~.
\end{array}
\end{equation}
Thus, we conclude that the Quantum Heisenberg-Weyl algebra corresponds to
the noncommutative space $SO_q(3)$ can be decoupled into three independent
subalgebras.   And then, one can investigate properties of the noncommutative
space by constructing Hilbert spaces of the three quantum subalgebras.

For the quantum algebra ${\cal A}_-$

\begin{equation}
\begin{array}{l}
{\cal D}_-{\cal X}^--{\cal X}^-{\cal D}_-=\mu_-^{-1}~,~~~~~~
\mu_- {\cal X}^-=q^2{\cal X}^-\mu_-~,\\
\mu_-{\cal D}_-=q^{-2}{\cal D}_-\mu_-~,~~~~~~\mu_-^{-1}\equiv
1+(q^{-2}-1){\cal X}^-{\cal D}_-~,
\end{array}
\end{equation}
we construct a quantum coherent state $\vert z\rangle_-$ as

\begin{equation}
\vert z\rangle_-=\exp_{q^2}(-\frac{1}{2}|q^{-2}z|)\sum_{m=0}^\infty
                   \frac{(-q^{-2}z)^m}{[m]_{q^2}!}({\cal D}_-)^m\vert 0\rangle_-~,
\end{equation}
where the notation of $q$-exponential
$$\exp_q(x)\equiv\sum_{n=0}^\infty \frac{x^n}{[n]_q!}~,~~~~
[n]_q!=[n]_q[n-1]_q\cdots[1]_q~,~~~~[n]_q=\frac{q^n-1}
{q-1}$$
has been used and the reference state
$\vert 0\rangle_-$ was chosen such that ${\cal X}^-\vert 0\rangle_- =0$.

As in the classical case, the coordinate ${\cal X}^-$ is diagonal in the
quantum coherent state basis

\begin{equation}
{\cal X}^-\vert z\rangle_- =z\vert z\rangle_-~.
\end{equation}
The real value of parameter $z$ may be interpreted as  position of an
indispersive wave-pocket\cite{11}. Here we should notice that $z$ can be
any complex number because
of us working on a general quantum orthogonal space. The complex values of
the quantum coherent state parameter are consistent with conjugate operation
on the noncommutative space.

Denote the quantum coherent state as

$$\vert 0,z\rangle_-\equiv \vert z\rangle_-~,$$
we can construct a representation for the quantum algebra ${\cal A}_-$ based on
the quantum coherent state as

\begin{equation}
\begin{array}{l}
{\cal X}^-\vert n,z\rangle_-=q^{2n}z\vert n,z\rangle_-~,\\
{\cal D}_-\vert n,z\rangle_-=-q^{-1-2n}\lambda^{-1}z^{-1}\vert n+1,z\rangle_-~,\\
\mu_-\vert n,z\rangle_-=\vert n-1,z\rangle_-~.
\end{array}
\end{equation}
It is clear from the Hilbert space representation of the quantum algebra
${\cal A}_-$ that the coordinates of the noncommutative orthogonal space is
discretely latticed with exponentially increasing space distances. In fact, this is in
agreement with the discrete
difference representation of quantum derivatives,

$${\cal D}f({\cal X})=\frac{f(q^2{\cal X})-f({\cal X})}{(q^2-1){\cal X}}~.$$
Similarly, we can construct the reference states $\vert 0,z\rangle_0$ and
$\vert 0,z\rangle_+$ as

\begin{equation}
\begin{array}{l}
\vert 0, z\rangle_0=\displaystyle\exp_{q^{-1}}(-\frac{1}{2}|qz|)\sum_{m=0}^\infty
                   \frac{(-qz)^m}{[m]_{q^{-1}}!}({\cal D}_0)^m\vert 0
                   \rangle_0~,~~~~~{\cal X}^0\vert 0\rangle_0=0~,\\
\vert 0, z\rangle_+=\displaystyle\exp_{q^{-2}}(-\frac{1}{2}|q^2z|)\sum_{m=0}^\infty
                   \frac{(-q^2z)^m}{[m]_{q^{-2}}!}({\cal D}_+)^m\vert 0
                   \rangle_+~,~~~~~{\cal X}^+\vert 0\rangle_+=0~.
\end{array}
\end{equation}
The corresponding representations of the quantum algebras ${\cal A}_0$

\begin{equation}
\begin{array}{l}
{\cal D}_0{\cal X}^0-{\cal X}^0{\cal D}_0=\mu_0^{1/2}~,~~~~~~
\mu_0 {\cal X}^0=q^2{\cal X}^0\mu_0~,\\
\mu_0{\cal D}_0=q^{-2}{\cal D}_0\mu_0~,~~~~~~\mu_0^{1/2}\equiv
1+(q-1){\cal X}^0{\cal D}_0~,
\end{array}
\end{equation}
and ${\cal A}_+$

\begin{equation}
\begin{array}{l}
{\cal D}_+{\cal X}^+-{\cal X}^+{\cal D}_+=\mu_+~,~~~~~~
\mu_+ {\cal X}^+=q^2{\cal X}^+\mu_+~,\\
\mu_+{\cal D}_+=q^{-2}{\cal D}_+\mu_+~,~~~~~~\mu_+\equiv
1+(q^2-1){\cal X}^+{\cal D}_+~,
\end{array}
\end{equation}
are of the forms

\begin{equation}
\begin{array}{l}
{\cal X}^0\vert n,z\rangle_0=q^{n}z\vert n,z\rangle_0~,\\
{\cal D}_0\vert n,z\rangle_0=\displaystyle\frac{q^{1-n}}{q-1}z^{-1}\vert n-1,z\rangle_0~,\\
\mu_0\vert n,z\rangle_0=\vert n-1,z\rangle_0~,
\end{array}
\end{equation}
and

\begin{equation}
\begin{array}{l}
{\cal X}^+\vert n,z\rangle_+=q^{2n}z\vert n,z\rangle_+~,\\
{\cal D}_+\vert n,z\rangle_+=q^{1-2n}\lambda^{-1}z^{-1}\vert n-1,z\rangle_+~,\\
\mu_+\vert n,z\rangle_+=\vert n-1,z\rangle_+~.
\end{array}
\end{equation}
It has been strongly argued that the most direct and obvious physical
cutoff of spacetime is
from the formation of microscopic black holes as soon as too much energy
would be accumulated into too small a region. Thus, from a physical point
of view, the black holes should provide for a natural cutoff all by
themselves. The cutoff distance scale is suspected to be the Planck
scale. Because of this origin of spacetime cutoff, any geometry we working
on there should be a same minimal cutoff $l_p$. For the classical geometry,
the spacetime regularization is equal distance and thus there is one degree
of freedom per Planck area. However, from the above discuss of
noncommutative geometry, the spacetime is discretely latticed with
exponentially increasing
space distances. Thus, much less information can be stored in the
noncommutative geometry. In fact, this may be the origin of the holography for
quantum system of gravity.

For the noncommutative space $SO_q(3)$ with radius $R$, if the assumed
minimal cutoff induced by the black holes themselves is the Planck scale
$l_p$, it is not difficult to count the degrees of freedom ${\cal N}_{\rm bulk}$,

\begin{equation}
\begin{array}{rcl}
{\cal N}_{\rm bulk}&\approx&\displaystyle\sum_{i=1}^N\frac{(q^{2i}l_p)^2}{(q^{2i}l_p-q^{2(i-1)}
l_p)^2}~,~~~~~~~~q^{2N}=R~,\\
&=&\displaystyle\frac{q^4\ln\left(\frac{R}{l_p}\right)}{2(q^2-1)^2\ln q}~.
\end{array}
\end{equation}
By taking the deformation parameter of the noncommutative space value to be
determined by the algebraic equation

\begin{equation}
q^{-4}(q^2-1)^2\ln q=\frac{\ln\left(\frac{R}{l_p}\right)}{8\pi\left(\frac{R}{l_p}
\right)^2}~,
\end{equation}
we can check that ${\cal N}_{\rm bulk}$ is exact equal to the
degrees of freedom on the classical limit of its boundary surface with radius
$R$, ${\cal N}_{\rm boundary}$,

\begin{equation}
{\cal N}_{\rm boundary}=\frac{4\pi R^2}{l_p^2}~.
\end{equation}
And then one can write down the entropy of our world at Planck scale,

\begin{equation}
S=4\pi M^2+C~,
\end{equation}
where $M$ is the mass of the world (black hole) in natural units and $C$ a
constant entropy can not determined.
This just is what was called holography or dimension reduction in quantum
theory of gravity by 't Hooft. At Planck scale, our world is not $3+1$
dimensional. Rather, the observable degrees of freedom can best be described
as if they were Boolean variables defined on a $2$ dimensional lattice,
evolving with time. It is clear now that the exact meaning of the
holography can interpreted as that the quantum theory of gravity in higher
dimensional noncommutative space is equivalent to the theory living on the
classical limit of spacetime boundary.
This supplies a reasonable picture for the 't Hooft's holography.

\section{Noncommutative AdS$_{2n}$ space and exponential regularization}

The noncommutative AdS$_{2n}$ space can be defined as the
$2n$-dimensional noncommutative real hyperboloid embedded in a ($2n+1$)-dimensional
space with coordinates $x^i$ ($i=-n,~-n+1,~\cdots,~-1,~0,~1,~\cdots,~n$),

\begin{equation}
\begin{array}{l}
\displaystyle\frac{1}{1+q^{2n-1}}C_{ij}x^ix^j=-\frac{1}{a^2}~,\\
{\overline x^i}=C_{ji}M_{jk}x^k~,
\end{array}
\end{equation}
where $\rho_{-i}=i-\frac{1}{2}$, $\rho_0=0$, $\rho_i=-i+\frac{1}{2}$, and
the metric $C_{ij}=q^{-\rho_i}\delta_{i,-j}$ and
$$M=\left(\begin{array}{ccc}
          1&  &  \\
           &\underbrace{\begin{array}{ccccc}
                      -1& & & &\\
                        &-1& & &\\
                        & &\ldots& &\\
                        & & &-1& \\
                        & & & &-1
                    \end{array}}_{2n-1}&\\
          &   & 1\end{array}\right)~.$$
It is easy to check that, at the $q\longrightarrow 1$ limit, the noncommutative
AdS$_{2n}$ space deduces to the familiar AdS$_{2n}$ space
$$\eta_{ab}x^ax^b=-\frac{1}{a^2}$$
in $R^{2n+1}$ with Cartesian coordinates $x^a$, where $\eta_{ab}=
{\rm diag}(-1,~\underbrace{1,~\cdots,~1}_{2n-1},~-1)$.

For the noncommutative AdS$_{2n}$ space, in components, the commutation relations
among coordinates are

\begin{equation}
\begin{array}{rcl}
x^ix^j&=&qx^jx^i~,~~~~~{\rm for}~i<j~ ~{\rm and}~~i\not=-j~,\\
x^ix^{-i}&=&x^{-i}x^i+\lambda q^{i-3/2}L_{i-1}\\
         &=&q^{-2}x^{-i}x^i+\lambda q^{i-3/2}L_i~,~~~~~{\rm for}~i>0~,
\end{array}
\end{equation}
where we have used the notation for intermediate lengths
$$L_i=\sum_{k=1}^i q^{\rho_k} x^{-k}x^k+\frac{q}{1+q}x^0x^0~.$$

By making use of the intermediate Laplacians
$$\Delta_i=\sum_{k=1}^iq^{\rho_k}\partial_k\partial_{-k}+\frac{q}{1+q}
\partial_0\partial_0~,$$
the algebra satisfied by the derivatives can be written compactly as

\begin{equation}
\begin{array}{rcl}
\partial_i\partial_j&=&q^{-1}\partial_j\partial_i~,~~~~~{\rm for}~i<j~ ~{\rm and}~~i\not=-j~,\\
\partial_{-i}\partial_i&=&\partial_i\partial_{-i}+\lambda q^{i-3/2}\Delta_{i-1}\\
                       &=&q^{-2}\partial_i\partial_{-i}+\lambda
                       q^{i-3/2}\Delta_i~,~~~~~{\rm for}~i>0~.
\end{array}
\end{equation}
The commutation relations among the coordinates and derivatives are as
follows

\begin{equation}
\begin{array}{l}
\partial_{-i}x^i=x^i\partial_{-i}~,~~~~~{\rm for}~i\not=0~,\\
\partial_ix^j=qx^j\partial_i~,~~~~~{\rm for}~j>-i,~{\rm and}~j\not=i~,\\
\partial_ix^j=qx^j\partial_i-q\lambda q^{-\rho_j-\rho_k}x^{-i}\partial_{-j}~,
~~~~~{\rm for}~j<-i~,~~{\rm and}~i\not=j~,\\
\partial_jx^j=1+q^2x^j\partial_j+q\lambda\displaystyle\sum_{k>j}x^k\partial_k
-q^{1-2\rho_j}\lambda x^{-j}\partial_{-j}~,~~~~~{\rm for}~j<0~,\\
\partial_0x^0=1+qx^0\partial_0+q\lambda\displaystyle\sum_{k>0}x^k\partial_k~,\\
\partial_jx^j=1+q^2x^j\partial_j+q\lambda\displaystyle\sum_{k>j}x^k\partial_k~,
~~~~~{\rm for}~j>0~.
\end{array}
\end{equation}
For convenient, we introduce the dilatation operator $\Lambda_m$ ($0<m\leq n$) as

$$
\displaystyle \Lambda_m=1+q\lambda E_m+q^{2m+1}\lambda^2L_m\Delta_m~,~~~~~
\displaystyle E_m=\sum_{j=-m}^mx^j\partial_j~.$$
The dilatation operator satisfies

\begin{equation}
\Lambda_m x^k=q^2x^k \Lambda_m~,~~~~~~~\Lambda_m\partial _k=q^{-2}\partial_k\Lambda_m~,~~~~{\rm for}~
\vert k\vert\leq m~.
\end{equation}
The noncommutative AdS$_{2n}$ space is accompanied with the
conjugation\cite{12}

\begin{equation}\label{ccc}
\begin{array}{l}
{\overline x^i}=C_{ji}M_{jk}x^k~,\\
{\overline \partial_i}=-q^{-2}\Lambda_{n}^{-1}C_{ij}M_{jk}[\Delta_n,x^k]~.
\end{array}
\end{equation}
For $k>0$, using the notations

$$\begin{array}{l}
y^{-k}=x^{-k}+q^{k+1/2}\lambda L_k\partial_{+k}~,\\
\delta_{-k}=\partial_{-k}+q^{k+1/2}\lambda \Delta_kx^{+k}~,
\end{array}$$
we can construct a set of independent basis\cite{13} on the noncommutative
AdS$_{2n}$ space,

\begin{equation}
\begin{array}{l}
{\cal X}^n=x^n~,\\
{\cal D}_n=\partial_n~,\\
{\cal X}^{+j}=\mu_{n}^{-1/2}\mu_{n-1}^{-1/2}\cdots
\mu_{j+1}^{-1/2}x^{+j}~,~~~~~{\rm for}~n>j\geq 0~,\\
{\cal D}_{+j}=\mu_{n}^{-1/2}\mu_{n-1}^{-1/2}\cdots
\mu_{j+1}^{-1/2}\partial_{+j}~,\\
{\cal X}^{-j}=\mu_{n}^{-1/2}\mu_{n-1}^{-1/2}\cdots\mu_{j+1}^{-1/2}
\Lambda_{+j}^{-1/2}\mu_{+j}^{-1/2}y^{-j}~,\\
{\cal D}_{-j}=q^{-1}\mu_{n}^{-1/2}\mu_{n-1}^{-1/2}\cdots\mu_{j+1}^{-1/2}
\Lambda_{+j}^{-1/2}\mu_{+j}^{-1/2}\delta_{-j}~,\\
{\cal X}^{-n}=\Lambda_{+n}^{-1/2}\mu_{+n}^{-1/2}y^{-n}~,\\
{\cal D}_{-j}=q^{-1}\Lambda_{+n}^{-1/2}\mu_{+n}^{-1/2}\delta_{-n}~,
\end{array}
\end{equation}
where $(\mu_{\pm i})^{\pm 1}={\cal D}_{\pm i}{\cal X}^{\pm i}-
{\cal X}^{\pm i}{\cal D}_{\pm i}$ and $\mu_0^{1/2}={\cal D}_0{\cal X}^0-
{\cal X}^0{\cal D}_0$.\\
We note that the $\mu_i$'s satisfy simple commutation relations with the
new variables ${\cal X}^j$ and ${\cal D}_j$,

$$[\mu_i,\mu_j]=0~,~~~~~~
\mu_i{\cal X}^j={\cal X}^j\mu_i\cdot\left\{\begin{array}{l}
                                           q^2~,~~~{\rm for}~i=j\\
                                           1~,~~~{\rm for}~i\not=j
                                           \end{array}\right.~,~~~~~~
\mu_i{\cal D}^j={\cal D}^j\mu_i\cdot\left\{\begin{array}{l}
                                           q^{-2}~,~~~{\rm for}~i=j\\
                                           1~,~~~{\rm for}~i\not=j
                                           \end{array}\right.~.$$

For the new basis of coordinates and derivatives on the noncommutative
AdS$_{2n}$ space, it is not difficult to show that

\begin{equation}
\begin{array}{l}
{\cal D}_{-k}{\cal X}^{-k}=1+q^{-2}{\cal X}^{-k}{\cal D}_{-k}~,~~~~~{\rm for}
~k>0~,\\
{\cal D}_{0}{\cal X}^{0}=1+q{\cal X}^{0}{\cal D}_{0}~,\\
{\cal D}_{+k}{\cal X}^{+k}=1+q^{2}{\cal X}^{+k}{\cal D}_{+k}~,\\[2mm]
[{\cal D}_i,{\cal D}_j]=0~,~~~~~[{\cal X}^i,{\cal X}^j]=0~,\\[2mm]
{\cal D}_i{\cal X}^j={\cal X}^j{\cal D}_i~,~~~~~{\rm for}~ i\not=j~.
\end{array}
\end{equation}
The noncommutative AdS$_{2n}$ space in terms of the ${\cal X}^i$ and ${\cal
D}_i$ is of the form

\begin{equation}
\sum_{j=1}^n q^{\rho_j-2(n-j)}\Lambda_j^{1/2}\mu_j^{-1/2}
{\cal X}^j{\cal X}^{-j}
+\frac{q^{-2n+1}}{1+q}{\cal X}^0{\cal X}^0=-\frac{1}{a^2}~.
\end{equation}
The conjugate operation on the independent set of operators ${\cal X}^j$ and
${\cal D}_j$ is induced from what on $x^i$ and $\partial_i$ (Eq.(\ref{ccc})),

\begin{equation}
\begin{array}{l}
{\overline {\cal X}^n}={\overline x^n}~,\\
{\overline {\cal D}_n}={\overline \partial_n}~,\\
{\overline {\cal X}^{+j}}={\overline x^{+j}}{\overline \mu_{j+1}}^{-1/2}
{\overline \mu_{j+2}}^{-1/2}\cdots
{\overline \mu_{n}}^{-1/2}~,~~~~~{\rm for}~n>j\geq 0~,\\
{\overline {\cal D}_{+j}}={\overline \partial_{+j}}{\overline \mu_{j+1}}^{-1/2}
{\overline \mu_{j+2}}^{-1/2}\cdots
{\overline \mu_{n}}^{-1/2}~,\\
{\overline {\cal X}^{-j}}={\overline y^{-j}}{\overline \mu_{+j}}^{-1/2}
{\overline \Lambda_{+j}}^{-1/2}{\overline \mu_{j+1}}^{-1/2}
{\overline \mu_{j+2}}^{-1/2}\cdots
{\overline \mu_{n}}^{-1/2}~,\\
{\overline {\cal D}_{-j}}=q^{-1}{\overline \delta_{-j}}{\overline
\mu_{+j}}^{-1/2}{\overline \Lambda_{+j}}^{-1/2}
{\overline \mu_{j+1}}^{-1/2}{\overline \mu_{j+2}}^{-1/2}\cdots
{\overline \mu_{n}}^{-1/2}~,\\
{\overline {\cal X}^{-n}}={\overline y^{-n}}{\overline \mu_{+n}}^{-1/2}
{\overline \Lambda_{+n}}^{-1/2}~,\\
{\overline {\cal D}_{-j}}=q^{-1}{\overline \delta_{-n}}
{\overline \mu_{+n}}^{-1/2}{\overline \Lambda_{+n}}^{-1/2}~.
\end{array}
\end{equation}
Thus, the Quantum Heisenberg-Weyl algebra corresponds to
the noncommutative AdS$_{2n}$ space can be decoupled into $(2n+1)$ independent
subalgebras.

For the quantum algebra ${\cal A}_{-k}$ ($ n\geq k>0$)

\begin{equation}
\begin{array}{l}
{\cal D}_{-k}{\cal X}^{-k}-{\cal X}^{-k}{\cal D}_{-k}=\mu_{-k}^{-1}~,~~~~~~
\mu_{-k} {\cal X}^{-k}=q^2{\cal X}^{-k}\mu_{-k}~,\\
\mu_-{\cal D}_{-k}=q^{-2}{\cal D}_{-k}\mu_{-k}~,~~~~~~\mu_-^{-1}\equiv
1+(q^{-2}-1){\cal X}^{-k}{\cal D}_{-k}~,
\end{array}
\end{equation}
we construct the quantum coherent state $\vert 0, z\rangle_{-k}$ as

\begin{equation}
\begin{array}{l}
\vert 0,z\rangle_{-k}=\displaystyle\exp_{q^2}(-\frac{1}{2}|q^{-2}z|)\sum_{m=0}^\infty
                   \frac{(-q^{-2}z)^m}{[m]_{q^2}!}({\cal D}_{-k})^m\vert
                   0\rangle_{-k}~,\\
{\cal X}^{-k}\vert 0,z\rangle_{-k}=z\vert 0,z\rangle_{-k}~,
\end{array}
\end{equation}
where the reference state
$\vert 0\rangle_{-k}$ was chosen such that ${\cal X}^{-k}\vert 0\rangle_{-k}
=0$.

From the coherent state $\vert 0,z\rangle_{-k}$,
we can construct a representation for the quantum algebra ${\cal A}_{-k}$ as

\begin{equation}
\begin{array}{l}
{\cal X}^{-k}\vert n,z\rangle_-=q^{2n}z\vert n,z\rangle_{-k}~,\\
{\cal D}_{-k}\vert n,z\rangle_-=-q^{-1-2n}\lambda^{-1}z^{-1}\vert n+1,
z\rangle_{-k}~,\\
\mu_{-k}\vert n,z\rangle_-=\vert n-1,z\rangle_{-k}~.
\end{array}
\end{equation}
The quantum coherent state corresponds to the quantum algebras ${\cal A}_0$

\begin{equation}
\begin{array}{l}
{\cal D}_0{\cal X}^0-{\cal X}^0{\cal D}_0=\mu_0^{1/2}~,~~~~~~
\mu_0 {\cal X}^0=q^2{\cal X}^0\mu_0~,\\
\mu_0{\cal D}_0=q^{-2}{\cal D}_0\mu_0~,~~~~~~\mu_0^{1/2}\equiv
1+(q-1){\cal X}^0{\cal D}_0~,
\end{array}
\end{equation}
is

\begin{equation}
\begin{array}{l}
\vert 0, z\rangle_0=\displaystyle\exp_{q^{-1}}(-\frac{1}{2}|qz|)\sum_{m=0}^\infty
                   \frac{(-qz)^m}{[m]_{q^{-1}}!}({\cal D}_0)^m\vert 0
                   \rangle_0~,~~~~~{\cal X}^0\vert 0\rangle_0=0~,\\
{\cal X}^{0}\vert 0,z\rangle_{0}=z\vert 0,z\rangle_{0}~.
\end{array}
\end{equation}
Representation of the quantum algebra ${\cal A}_0$ based on the quantum
coherent state $\vert 0,z\rangle_0$ is as

\begin{equation}
\begin{array}{l}
{\cal X}^0\vert n,z\rangle_0=q^{n}z\vert n,z\rangle_0~,\\
{\cal D}_0\vert n,z\rangle_0=\displaystyle\frac{q^{1-n}}{q-1}z^{-1}
\vert n-1,z\rangle_0~,\\
\mu_0\vert n,z\rangle_0=\vert n-1,z\rangle_0~.
\end{array}
\end{equation}
The quantum coherent state corresponds to the quantum algebra
${\cal A}_{+k}$

\begin{equation}
\begin{array}{l}
{\cal D}_{+k}{\cal X}^{+k}-{\cal X}^{+k}{\cal D}_{+k}=\mu_{+k}~,~~~~~~
\mu_{+k} {\cal X}^{+k}=q^2{\cal X}^{+k}\mu_{+k}~,\\
\mu_{+k}{\cal D}_{+k}=q^{-2}{\cal D}_{+k}\mu_{+k}~,~~~~~~\mu_{+k}\equiv
1+(q^2-1){\cal X}^{+k}{\cal D}_{+k}~,
\end{array}
\end{equation}
is of the form

\begin{equation}
\begin{array}{l}
\vert 0, z\rangle_{+k}=\displaystyle\exp_{q^{-2}}(-\frac{1}{2}|q^2z|)\sum_{m=0}^\infty
                   \frac{(-q^2z)^m}{[m]_{q^{-2}}!}({\cal D}_{+k})^m\vert 0
                   \rangle_{+k}~,~~~~~{\cal X}^{+k}\vert 0\rangle_{+k}=0~,\\
{\cal X}^{+k}\vert 0,z\rangle_{+k}=z\vert 0,z\rangle_{+k}~.
\end{array}
\end{equation}
Then, by writing down the representation of the quantum algebra
${\cal A}_{+k}$

\begin{equation}
\begin{array}{l}
{\cal X}^{+k}\vert n,z\rangle_{+k}=q^{2n}z\vert n,z\rangle_{+k}~,\\
{\cal D}_{+k}\vert n,z\rangle_{+k}=q^{1-2n}\lambda^{-1}z^{-1}\vert n-1,z\rangle_{+k}~,\\
\Lambda_{+k}\vert n,z\rangle_{+k}=\vert n-1,z\rangle_{+k}~,
\end{array}
\end{equation}
we complete the Hilbert space constructure of the noncommutative AdS$_{2n}$
space.
This shows that the noncommutative AdS$_{2n}$ space is discretely latticed with
exponentially
increasing space distances. The minimal cutoff induced by the quantum
gravity itself is the Planck scale $l_p$. As in the case of noncommutative
orthogonal
space, the exponential regularization may effectively reduce the amount
of observable degrees of freedom of the noncommutative AdS$_{2n}$ space, even one can
not enumerate it exactly because of being infinite\cite{14}. An adequately
adopted noncommutative
deformation parameter $q$ (it may be even more closer to $1$ than the case
of limited geometry) can equate the entropy of the quantum system of
gravity in the bulk of noncommutative AdS$_{2n}$ space and what on the classical
limit of its boundary. The commutative boundary is equal distance regularized
and possesses conformally invariant symmetry. This is crucial for the
AdS/CFT correspondence.

\section{Noncommutative AdS$_{2n-1}$ space with exponential regularization}

For the noncommutative AdS$_{2n-1}$ space, commutation relations among
coordinates $x^i$ ($i=-n,~-n+1,~\cdots,~-2,~-1,~+1,~+2,~\cdots,~n$) are

\begin{equation}
\begin{array}{rcl}
x^ix^j&=&qx^jx^i~,~~~~~{\rm for}~i<j~ ~{\rm and}~~i\not=-j~,\\
x^ix^{-i}&=&x^{-i}x^i+\lambda q^{i-2}L_{i-1}\\
         &=&q^{-2}x^{-i}x^i+\lambda q^{i-2}L_i~,~~~~~{\rm for}~i>0~,
\end{array}
\end{equation}
where we have used the notation for intermediate lengths
$$L_i=\sum_{k=1}^i q^{\rho_k} x^{-k}x^k~,~~~~\rho_{-k}=k-1~,~~~~\rho_k=-k+1~.$$

By making use of the intermediate Laplacians
$$\Delta_i=\sum_{k=1}^iq^{\rho_k}\partial_k\partial_{-k},$$
the algebra satisfied by the derivatives can be written compactly as

\begin{equation}
\begin{array}{rcl}
\partial_i\partial_j&=&q^{-1}\partial_j\partial_i~,~~~~~{\rm for}~i<j~ ~{\rm and}~~i\not=-j~,\\
\partial_{-i}\partial_i&=&\partial_i\partial_{-i}+\lambda q^{i-2}\Delta_{i-1}\\
                       &=&q^{-2}\partial_i\partial_{-i}+\lambda
                       q^{i-2}\Delta_i~,~~~~~{\rm for}~i>0~.
\end{array}
\end{equation}
The commutation relations among the coordinates and derivatives are as
follows

\begin{equation}
\begin{array}{l}
\partial_{-i}x^i=x^i\partial_{-i}~,~~~~~{\rm for}~i\not=0~,\\
\partial_ix^j=qx^j\partial_i~,~~~~~{\rm for}~j>-i,~{\rm and}~j\not=i~,\\
\partial_ix^j=qx^j\partial_i-q\lambda q^{-\rho_j-\rho_k}x^{-i}\partial_{-j}~,
~~~~~{\rm for}~j<-i~,~~{\rm and}~i\not=j~,\\
\partial_jx^j=1+q^2x^j\partial_j+q\lambda\displaystyle\sum_{k>j}x^k\partial_k
-q^{1-2\rho_j}\lambda x^{-j}\partial_{-j}~,~~~~~{\rm for}~j<0~,\\
\partial_jx^j=1+q^2x^j\partial_j+q\lambda\displaystyle\sum_{k>j}x^k\partial_k~,
~~~~~{\rm for}~j>0~.
\end{array}
\end{equation}
For convenient, we introduce the dilatation operator $\Lambda_m$ ($0<m\leq n$) as

$$\Lambda_m=1+q\lambda E_m+q^{2m}\lambda^2(1+q^{2n-2})L_m\Delta_m~,~~~~~
E_m=\sum_{j=-m}^mx^j\partial_j~.$$
The dilatation operator satisfies

\begin{equation}
\Lambda_m x^k=q^2x^k \Lambda_m~,~~~~~~~\Lambda_m\partial _k=q^{-2}\partial_k\Lambda_m~,~~~~{\rm for}~
<0\vert k\vert\leq m~.
\end{equation}
The noncommutative AdS$_{2n-1}$ space is accompanied with the conjugation

\begin{equation}
\begin{array}{l}
{\overline x^i}=C_{ji}N_{jk}x^k~,\\
{\overline \partial_i}=-q^{-2}\Lambda_{n}^{-1}C_{ij}N_{jk}[\Delta_n,x^k]~,
\end{array}
\end{equation}
where
$$N=\left(\begin{array}{ccc}
          1&  &  \\
           &\underbrace{\begin{array}{ccccc}
                      -1& & & &\\
                        &-1& & &\\
                        & &\ldots& &\\
                        & & &-1& \\
                        & & & &-1
                    \end{array}}_{2n-2}&\\
          &   & 1\end{array}\right)~.$$
For $k>0$, using the notations

$$\begin{array}{l}
y^{-k}=x^{-k}+q^{k+1/2}\lambda (1+q^{2n-2})L_k\partial_{+k}~,\\
\delta_{-k}=\partial_{-k}+q^{k+1/2}\lambda \Delta_kx^{+k}~,
\end{array}$$
we can construct a new set of variables for the coordinates and derivatives,

\begin{equation}
\begin{array}{l}
{\cal X}^n=x^n~,\\
{\cal D}_n=\partial_n~,\\
{\cal X}^{+j}=\mu_{n}^{-1/2}\mu_{n-1}^{-1/2}\cdots
\mu_{j+1}^{-1/2}x^{+j}~,~~~~~{\rm for}~n>j> 0~,\\
{\cal D}_{+j}=\mu_{n}^{-1/2}\mu_{n-1}^{-1/2}\cdots
\mu_{j+1}^{-1/2}\partial_{+j}~,\\
{\cal X}^{-j}=\mu_{n}^{-1/2}\mu_{n-1}^{-1/2}\cdots\mu_{j+1}^{-1/2}
\Lambda_{+j}^{-1/2}\mu_{+j}^{-1/2}y^{-j}~,\\
{\cal D}_{-j}=q^{-1}\mu_{n}^{-1/2}\mu_{n-1}^{-1/2}\cdots\mu_{j+1}^{-1/2}
\Lambda_{+j}^{-1/2}\mu_{+j}^{-1/2}\delta_{-j}~,\\
{\cal X}^{-n}=\Lambda_{+n}^{-1/2}\mu_{+n}^{-1/2}y^{-n}~,\\
{\cal D}_{-j}=q^{-1}\Lambda_{+n}^{-1/2}\mu_{+n}^{-1/2}\delta_{-n}~,
\end{array}
\end{equation}
where $(\mu_{\pm i})^{\pm 1}={\cal D}_{\pm i}{\cal X}^{\pm i}-
{\cal X}^{\pm i}{\cal D}_{\pm i}$.\\
Commutation relations among the new basis of coordinates and
derivatives on the noncommutative AdS$_{2n-1}$ space are

\begin{equation}
\begin{array}{l}
{\cal D}_{-k}{\cal X}^{-k}=1+q^{-2}{\cal X}^{-k}{\cal D}_{-k}~,~~~~~{\rm for}
~k>0~,\\
{\cal D}_{+k}{\cal X}^{+k}=1+q^{2}{\cal X}^{+k}{\cal D}_{+k}~,\\[2mm]
[{\cal D}_i,{\cal D}_j]=0~,~~~~~[{\cal X}^i,{\cal X}^j]=0~,\\[2mm]
{\cal D}_i{\cal X}^j={\cal X}^j{\cal D}_i~,~~~~~{\rm for}~ i\not=j~.
\end{array}
\end{equation}
The noncommutative AdS$_{2n-1}$ space in terms of the ${\cal X}^i$ and ${\cal
D}_i$ is of the form

\begin{equation}
\sum_{j=1}^n q^{\rho_j-2(n-j)}\Lambda_j^{1/2}\mu_j^{-1/2}
{\cal X}^j{\cal X}^{-j}=-\frac{1}{a^2}~.
\end{equation}
The conjugate operation on the operators ${\cal X}^j$ and
${\cal D}_j$ is induced by the conjugation on $x^i$ and $\partial_i$,

\begin{equation}
\begin{array}{l}
{\overline {\cal X}^n}={\overline x^n}~,\\
{\overline {\cal D}_n}={\overline \partial_n}~,\\
{\overline {\cal X}^{+j}}={\overline x^{+j}}{\overline \mu_{j+1}}^{-1/2}
{\overline \mu_{j+2}}^{-1/2}\cdots
{\overline \mu_{n}}^{-1/2}~,~~~~~{\rm for}~n>j> 0~,\\
{\overline {\cal D}_{+j}}={\overline \partial_{+j}}{\overline \mu_{j+1}}^{-1/2}
{\overline \mu_{j+2}}^{-1/2}\cdots
{\overline \mu_{n}}^{-1/2}~,\\
{\overline {\cal X}^{-j}}={\overline y^{-j}}{\overline \mu_{+j}}^{-1/2}
{\overline \Lambda_{+j}}^{-1/2}{\overline \mu_{j+1}}^{-1/2}
{\overline \mu_{j+2}}^{-1/2}\cdots
{\overline \mu_{n}}^{-1/2}~,\\
{\overline {\cal D}_{-j}}=q^{-1}{\overline \delta_{-j}}{\overline
\mu_{+j}}^{-1/2}{\overline \Lambda_{+j}}^{-1/2}
{\overline \mu_{j+1}}^{-1/2}{\overline \mu_{j+2}}^{-1/2}\cdots
{\overline \mu_{n}}^{-1/2}~,\\
{\overline {\cal X}^{-n}}={\overline y^{-n}}{\overline \mu_{+n}}^{-1/2}
{\overline \Lambda_{+n}}^{-1/2}~,\\
{\overline {\cal D}_{-j}}=q^{-1}{\overline \delta_{-n}}
{\overline \mu_{+n}}^{-1/2}{\overline \Lambda_{+n}}^{-1/2}~.
\end{array}
\end{equation}
Then, the Quantum Heisenberg-Weyl algebra corresponds to
the noncommutative AdS$_{2n-1}$ space is decoupled into $2n$ independent
subalgebras.

For the quantum algebra ${\cal A}_{-k}$ ($ n\geq k>0$)

\begin{equation}
\begin{array}{l}
{\cal D}_{-k}{\cal X}^{-k}-{\cal X}^{-k}{\cal D}_{-k}=\mu_{-k}^{-1}~,~~~~~~
\mu_{-k} {\cal X}^{-k}=q^2{\cal X}^{-k}\mu_{-k}~,\\
\mu_-{\cal D}_{-k}=q^{-2}{\cal D}_{-k}\mu_{-k}~,~~~~~~\mu_-^{-1}\equiv
1+(q^{-2}-1){\cal X}^{-k}{\cal D}_{-k}~,
\end{array}
\end{equation}
we have the quantum coherent state $\vert 0,z\rangle_{-k}$ as

\begin{equation}
\begin{array}{l}
\vert 0,z\rangle_{-k}=\displaystyle\exp_{q^2}(-\frac{1}{2}|q^{-2}z|)\sum_{m=0}^\infty
                   \frac{(-q^{-2}z)^m}{[m]_{q^2}!}({\cal D}_{-k})^m\vert
                   0\rangle_{-k}~,~~~~~{\cal X}^{-k}\vert 0\rangle_{-k}
                  =0~,\\
{\cal X}^{-k}\vert 0,z\rangle_{-k} =z\vert 0,z\rangle_{-k}~.
\end{array}
\end{equation}
From the coherent state $\vert 0,z\rangle_{-k}$,
we can construct a representation for the quantum algebra ${\cal A}_{-k}$ as

\begin{equation}
\begin{array}{l}
{\cal X}^{-k}\vert n,z\rangle_-=q^{2n}z\vert n,z\rangle_{-k}~,\\
{\cal D}_{-k}\vert n,z\rangle_-=-q^{-1-2n}\lambda^{-1}z^{-1}\vert n+1,
z\rangle_{-k}~,\\
\mu_{-k}\vert n,z\rangle_-=\vert n-1,z\rangle_{-k}~.
\end{array}
\end{equation}
Similarly, we can construct reference state$\vert 0,z\rangle_{+k}$
($n\geq k>0$) as

\begin{equation}
\begin{array}{l}
\vert 0, z\rangle_{+k}=\displaystyle\exp_{q^{-2}}(-\frac{1}{2}|q^2z|)\sum_{m=0}^\infty
                   \frac{(-q^2z)^m}{[m]_{q^{-2}}!}({\cal D}_{+k})^m\vert 0
                   \rangle_{+k}~,~~~~~{\cal X}^{+k}\vert 0\rangle_{+k}=0~,\\
{\cal X}^{+k}\vert 0,z\rangle_{+k} =z\vert 0,z\rangle_{+k}~.
\end{array}
\end{equation}
The corresponding representation of the quantum algebras ${\cal A}_{+k}$

\begin{equation}
\begin{array}{l}
{\cal D}_{+k}{\cal X}^{+k}-{\cal X}^{+k}{\cal D}_{+k}=\mu_{+k}~,~~~~~~
\mu_{+k} {\cal X}^{+k}=q^2{\cal X}^{+k}\mu_{+k}~,\\
\mu_{+k}{\cal D}_{+k}=q^{-2}{\cal D}_{+k}\mu_{+k}~,~~~~~~\mu_{+k}\equiv
1+(q^2-1){\cal X}^{+k}{\cal D}_{+k}~,
\end{array}
\end{equation}
is of the form

\begin{equation}
\begin{array}{l}
{\cal X}^{+k}\vert n,z\rangle_{+k}=q^{2n}z\vert n,z\rangle_{+k}~,\\
{\cal D}_{+k}\vert n,z\rangle_{+k}=q^{1-2n}\lambda^{-1}z^{-1}\vert n-1,z\rangle_{+k}~,\\
\Lambda_{+k}\vert n,z\rangle_{+k}=\vert n-1,z\rangle_{+k}~.
\end{array}
\end{equation}
As the noncommutative AdS$_{2n}$ space did, the noncommutative AdS$_{2n-1}$
space is also discretely latticed with exponentially
increasing space distances. The minimal cutoff induced by the quantum
gravity itself is the Planck scale $l_p$. The exponential regularization
effectively reduces degrees of freedom in the noncommutative AdS$_{2n-1}$ space.
A very small amount of displacement of the noncommutative deformation parameter $q$
from unity equates the entropy of the quantum theory of
gravity in the bulk of noncommutative AdS$_{2n-1}$ space and what on the commutative
limit of its boundary surface. The commutative boundary is equal distance
regularized and possesses a conformally invariant symmetry.
Thus, the equivalent theory living on the spacetime boundary of the quantum
system of gravity
on the background of noncommutative AdS space is a conformal field theory. This
is the bases for the AdS/CFT correspondence.

\section{Concluding remarks}

In this paper, by constructing Hilbert space with quantum coherent state as
reference one, we have presented a kind of special regularization with
exponentially increasing spacetime cutoff for both orthogonal and AdS space
based on noncommutative geometry. It has been used to argue that the same
minimal cutoff of any geometry is the Planck scale $l_p$, that the most
direct and obvious physical cutoff is from the formation of microscopic
black holes as soon as too much energy would be accumulated
into too small a region. We have obtained results which show a very small
($<10^{-15}$) displacement of the noncommutative deformation parameter from
its classical value ($1$) reduces sharply the entropy of quantum system of
gravity.
The noncommutative deformation parameter was
determined by a well-defined algebraic equation. The noncommutative space
$SO_q(3)$ with such a
deformation parameter have the same entropy or degrees of freedom with classical
$S^3$ surface. This is just the so called holography for quantum theory of
gravity. The holography makes the
quantum theory of gravity on the noncommutative AdS$_{d}$ space equivalent
to the conformally invariant quantum field theory living
on the classical limit of its boundary. This is the bases of the
AdS/CFT correspondence of string theory and M theory. Here we should stress
that the proper geometry for quantum gravity may be noncommutative but not
classical continuum geometry. This is in agreement with the long time
suspecting that the small scale structure of spacetime might not be
adequately described by classical continuum geometry and the noncommutative
spacetime might be a realistic picture of how spacetime behaves near
the Planck scale.  Strong quantum fluctuations of gravity at this spacetime
scale may make points in space fuzzy.
All of the strangeness'  for the quantum theories in different spacetime
dimensions ever could be equivalent are come from the noncommutative geometry
description of the quantum gravity. The exact form of the AdS/CFT is
concerned with noncommutative AdS space and the classical limit of its boundary
surface (on which the conformal field theory lives).
This suggests that the gravity-gauge theory connection should be of the form:
the string theory or M theory on the noncommutative background of the form
AdS$_d^q\times M^q_{D-d}$ is dual to a conformal
field theory living on the classical limit of spacetime boundary.
For the Type IIB string theory on the noncommutative
background AdS$^q_5\times S_q^5$ spectra can compare with low-order
correlation functions of the $3+1$ dimensional ${\cal N}=4$ $SU(N)$ super
Yang-Mills theory. The dual super
Yang-Mills theory lives on the classical limit of noncommutative AdS
spacetime boundary. Only both in the large $N$ and commutative limit,
the string theory is weakly coupled and the supergravity is a good
approximation to it. Therefore, in fact as 't Hooft suspected\cite{01},
the nature is much more crazy at the Planck scale than even string theorists
could have imagined. Formalisms of string theory (gravity) on noncommutative
geometry have to be constructed to gain sights of unification of gravity and
quantum mechanics.

\bigskip
\bigskip
\centerline{\large\bf Acknowledgments }

I would like to thank Prof. J. Wess for introducing the problem to me and
for enlightening discussions.
I am grateful to H. Steinacker for valuable discussions.
The work was supported in part by the National Science Foundation of China
under Grant 19625512.


\begin{thebibliography}{99}
\bibitem{01} G. 't Hooft, ``Dimension reduction in quantum gravity'', in
             {\it Salamfest 1993}, P284, gr-qc/9310026.
\bibitem{02} L. Susskind, ``The world as a hologram'', J. Math. Phys.
             {\bf 36} (1995) 6377, hep-th/9409089.
\bibitem{03} J. Maldacena, ``The large $N$ limit of superconformal field
             theories and supergravity'', Adv. Theor. Math. Phys. {\bf 2}
             (1998) 231, hep-th/9711200.
\bibitem{04} S. Gubser, I. Klebanov and A. Polyakov, ``Gauge theory
             correlators from non-critical string theory'', Phys. Lett.
             B{\bf 428} (1998) 105, hep-th/9802109.
\bibitem{05} E. Witten, ``Anti de Sitter space and holography'', Adv. Theor.
             Math. Phys. {\bf 2} (1998) 253, hep-th/9802150.
\bibitem{06} J. Wess, ``$q$-defomed phase space and its lattice structure'',
             Int. J. Mod. Phys. {\bf 12} (1997) 4997.
\bibitem{07} J. Madore, ``Gravity on fuzzy space-time'', ESI preprint 478 (1997),
             gr-qc/9709002.
\bibitem{08} J. Wess and B. Zumino, ``Covariant differential calculus on the
             quantum hyperplane'', Nucl. Phys. (Proc. Suppl.) {\bf 18}B (1990)
             302.
\bibitem{09} A. Connes, ``Noncommutative Geometry'', Academic Press (1994).
\bibitem{10} L. D. Faddeev, N. Yu. Reshetikhin and L. A. Takhtajan,
             ``Quantization of Lie groups and Lie algebras'',
             Leningrad Math. J. {\bf 1} (1990) 193.
\bibitem{11} J. Klauder and B. Skagerstam, ``Coherent states, applications
             in physics and mathematical physics'', World Scientific (1985).
\bibitem{12} Z. Chang, ``Quantum Anti-de Sitter space'', MPI-PhT 99-15,
             hep-th/9904091.
\bibitem{13} O. Ogievetsky, ``Differential operators on quantum spaces for
             $GL_q(n)$ and $SO_q(n)$'', Lett. Math. Phys. {\bf 24} (1992) 245.
\bibitem{14} L. Susskind and E. Witten, ``The holographic bound in Anti-de
             Sitter space'', SU-ITP-98-39, IASSNS-98-44, hep-th/9805114.
\end{thebibliography}
\end{document}